\documentclass[twocolumn,prl,floatfix,superscriptaddress,nobibnotes,nolongbibliography]{revtex4-2}

\usepackage{amsmath}
\usepackage{amssymb}
\usepackage{graphicx}
\usepackage{hyperref}
\usepackage[normalem]{ulem}

\usepackage{xcolor}

\hypersetup{
    colorlinks=true,
    linkcolor={blue!80!black},
    citecolor={blue!80!black},
    urlcolor={blue!80!black}
}

\usepackage{color}
 
\begin{document}

\title{Magnetic-field response of generalized Wigner crystals in twisted MoTe$_2$}
\author{F. K. Wolff}
\affiliation{Niels Bohr Institute, University of Copenhagen, 2100 Copenhagen, Denmark}
\author{S. Jiao}
\affiliation{Niels Bohr Institute, University of Copenhagen, 2100 Copenhagen, Denmark}
\author{H. Park}
\affiliation{Department of Physics, University of Washington, Seattle, Washington 98195, USA}
\author{J. Cai}
\affiliation{Department of Physics, University of Washington, Seattle, Washington 98195, USA}
\author{E. Anderson}
\affiliation{Department of Physics, University of Washington, Seattle, Washington 98195, USA}
\author{X. Liu}
\affiliation{Department of Materials Science and Engineering, University of Washington, Seattle, Washington 98195, USA}
\author{C. Wang}
\affiliation{Department of Materials Science and Engineering, University of Washington, Seattle, Washington 98195, USA}
\author{T.~Taniguchi}
\affiliation{Research Center for Materials Nanoarchitectonics, National Institute for Materials Science, 1-1 Namiki, Tsukuba 305-0044, Japan}
\author{K.~Watanabe}
\affiliation{Research Center for Electronic and Optical Materials, National Institute for Materials Science, 1-1 Namiki, Tsukuba 305-0044, Japan}
\author{X. Xu}
\affiliation{Department of Physics, University of Washington, Seattle, Washington 98195, USA}
\affiliation{Department of Materials Science and Engineering, University of Washington, Seattle, Washington 98195, USA}
\author{D. Xiao}
\affiliation{Department of Materials Science and Engineering, University of Washington, Seattle, Washington 98195, USA}
\affiliation{Department of Physics, University of Washington, Seattle, Washington 98195, USA}
\author{S. Vaitiek\.{e}nas}
\affiliation{Niels Bohr Institute, University of Copenhagen, 2100 Copenhagen, Denmark}

\date{\today}

\begin{abstract}
We report a magnetotransport study of fractional hole-filling states in the layer-polarized regime of a twisted MoTe$_2$ homobilayer. 
A perpendicular magnetic field suppresses the resistance peak at the two-thirds filling while enhancing the one-third peak by several orders of magnitude. 
Both remain near their commensurate densities. 
We argue that these commensurate states are generalized Wigner crystals whose distinct charge configurations define different magnetic lattices, leading to contrasting responses to spin polarization. 
Near half filling, we observe a qualitatively different state whose resistance maximum shifts approximately quadratically with field and exhibits pronounced transport anisotropy consistent with a stripe phase. These results reveal a strong dependence of magnetotransport on fractional filling and open a route to magnetic control of transport in moiré charge-ordered states.
\end{abstract}

\maketitle

Generalized Wigner crystals form a diverse family of charge-ordered states stabilized by long-range Coulomb interactions at fractional filling of a moiré lattice~\cite{Wu2018, Padhi2021}. 
Following their discovery in WSe$_2$/WS$_2$ heterobilayers, experiments have identified a broad sequence of fractional states and resolved triangular, honeycomb, and stripe charge patterns~\cite{Regan2020, Xu2020, Huang2021, Li2021, Jin2021}. 
The localized carriers retain magnetic moments, so each charge configuration also defines a magnetic lattice~\cite{Pan2020, MoralesDuran2023}. 
Changing the arrangement of occupied sites reorganizes the exchange pathways and magnetic frustration, allowing distinct magnetic states to emerge at different fractional fillings~\cite{MoralesDuran2022}. In transition-metal dichalcogenide moiré systems, strong spin–orbit coupling further locks spin and valley degrees of freedom~\cite{Jin2018}, which can strongly constrain their magnetic response and lead to topological phases~\cite{Wu2019}, including recently observed fractional Chern insulators~\cite{Cai2023, Park2023, Zeng2023, Xu2023}.

Magneto-optical measurements have established antiferromagnetic exchange in moiré Mott states~\cite{Tang2020} and competing interactions in the two-thirds Wigner–Mott state~\cite{Tang2023}. 
Transport experiments have also demonstrated magnetic-field control of correlated states in graphene and semiconductor moiré systems~\cite{Chen2023, Zong2025, Chang2026}. 
An external field couples to the magnetic moments and competes with their exchange interactions, with consequences for both the stability of charge order and the energy or mobility of charged excitations~\cite{MoralesDuran2023, Slagle2020}. 
Magnetotransport therefore provides a natural way to explore how the magnetic response of different charge-ordered states feeds back on charge motion.

\begin{figure}[h!]
\includegraphics[width=1\linewidth]{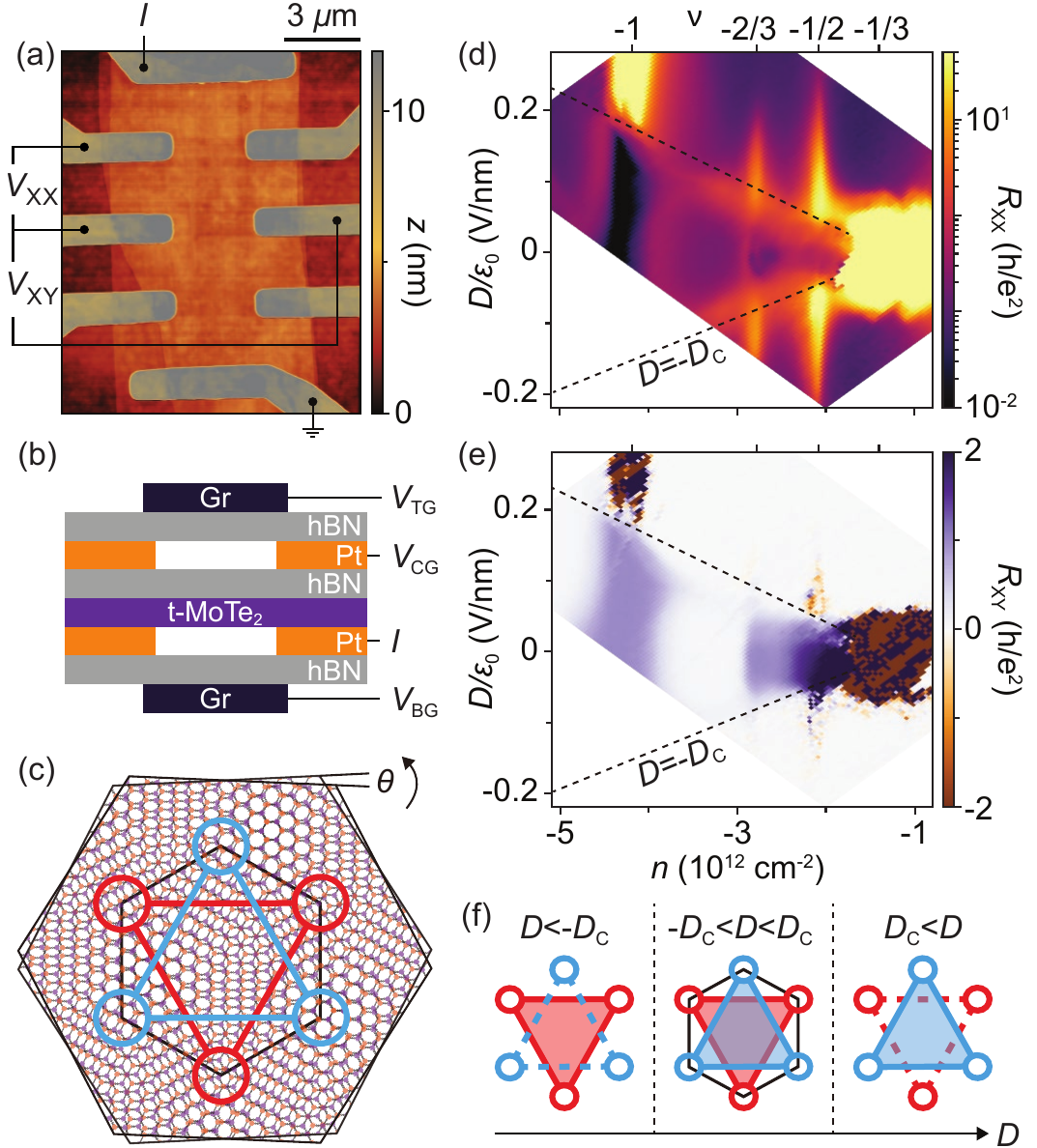}
\caption{
(a)~Atomic-force micrograph of the device overlaid with the measurement setup. 
(b)~Schematic cross section showing the twisted MoTe$_2$ bilayer, Pt contacts, and the three-gate geometry. The top and bottom gates at voltages $V_{\rm TG}$ and $V_{\rm BG}$ are used to tune the charge-carrier density and the perpendicular displacement field, whereas the contact gate at $V_{\rm CG}$ improves the contact transparency. 
(c) Schematic illustration of the moiré superlattice formed in twisted bilayer MoTe$_2$, highlighting the two moiré sublattices. 
(d) and (e) Longitudinal and transverse resistance, $R_{\rm XX}$ and $R_{\rm XY}$, measured as a function of carrier density, $n$, and displacement field, $D$ at $\vert\mu_0H\vert=0.2$~T. The dashed lines $\vert D\vert=D_{\rm C}$ mark the approximate boundary between the layer-hybridized ($\vert D\vert<D_{\rm C}$) and layer-polarized ($\vert D\vert>D_{\rm C}$) regimes. 
(f) Pictogram illustrating the evolution of the relative sublattice occupation with increasing displacement field.
}
\label{fig:1}
\end{figure}

Here, we investigate the magnetotransport of fractional states in hole-doped twisted MoTe$_2$ at large displacement field, where the system is driven into the layer-polarized triangular-lattice limit. 
A perpendicular magnetic field suppresses the resistance peak at hole filling $\nu=-2/3$, driving the transport from insulator-like toward metal-like behavior, whereas at $\nu=-1/3$ it produces a sharp resistance peak that grows by several orders of magnitude.
Both responses develop on a field scale of order 1~T, while the resistance maxima remain close to their commensurate densities.
We interpret this contrasting response in terms of generalized Wigner crystals with distinct magnetic lattices set by their charge configurations.
Near $\nu=-1/2$, we find a qualitatively different state whose resistance maximum shifts approximately quadratically with field and exhibits pronounced transport anisotropy, compatible with a stripe-like charge order.
These findings establish twisted MoTe$_2$ as a platform for exploring the interplay between magnetism and charge order, opening a route to magnetic control of transport across distinct correlated states.

The measured device consists of a twisted MoTe$_2$ homobilayer with a twist angle of approximately $3.9^\circ$, encapsulated in hexagonal boron nitride with graphite top and bottom gates, patterned into a Hall bar geometry [Figs.~\ref{fig:1}(a) and \ref{fig:1}(b)] following fabrication methods reported in Refs.~\cite{Cai2023, Park2023}. 
The dual-gate geometry [Fig.~\ref{fig:1}(b)] enables independent control of carrier density, $n$, and perpendicular displacement field, $D$.
Negative filling $\nu$ denotes hole doping, with $|\nu|$ giving the number of holes per moiré unit cell.
A local Pt contact gate increased the hole density near the contacts. 
All measurements were carried out using standard low-frequency lock-in techniques in a dilution refrigerator with a base temperature of 20~mK, and the perpendicular magnetic field, $H$, was applied using a superconducting vector magnet.
Longitudinal (transverse) resistance, $R_{\rm XX}$ ($R_{\rm XY}$), was symmetrized (anti-symmetrized) under reversal of $H$ to account for contact misalignment~\cite{Look1989}. 

Stacking two MoTe$_2$ monolayers with a small twist angle results in a moiré superlattice with two inequivalent sublattices; see Fig.~\ref{fig:1}(c).  
At small $\vert D\vert$, orbitals associated with opposite layers hybridize into a honeycomb moiré lattice~\cite{Wu2019, Zhang2021}, the regime in which integer and fractional anomalous Hall states have been reported~\cite{Cai2023, Park2023, Zeng2023}.
Increasing $\vert D\vert$ lifts the degeneracy of the two sublattices and favors occupation of one of them, driving the system toward a triangular-lattice limit~\cite{Anderson2023}.
At $\nu=-2/3$, this tuning has previously been shown to drive a transition from a fractional Chern insulator to a topologically trivial correlated state~\cite{Xu2023}. 
In our device, this crossover is visible in $R_{\rm XX}$ and $R_{\rm XY}$ as a function of $n$ and $D$, which reveal two regimes separated by a wedge-shaped boundary at $\vert D\vert = D_{\rm C}$; see Figs.~\ref{fig:1}(d) and \ref{fig:1}(e). 
The layer-polarized assignment refers to the dominant orbital character, illustrated schematically in Fig.~\ref{fig:1}(f).
We focus below on the nominally layer-polarized regime, $|D|>D_{\rm C}$, where the carriers predominantly occupy a single low-energy sublattice.

\begin{figure}[t]
\includegraphics[width=1\linewidth]{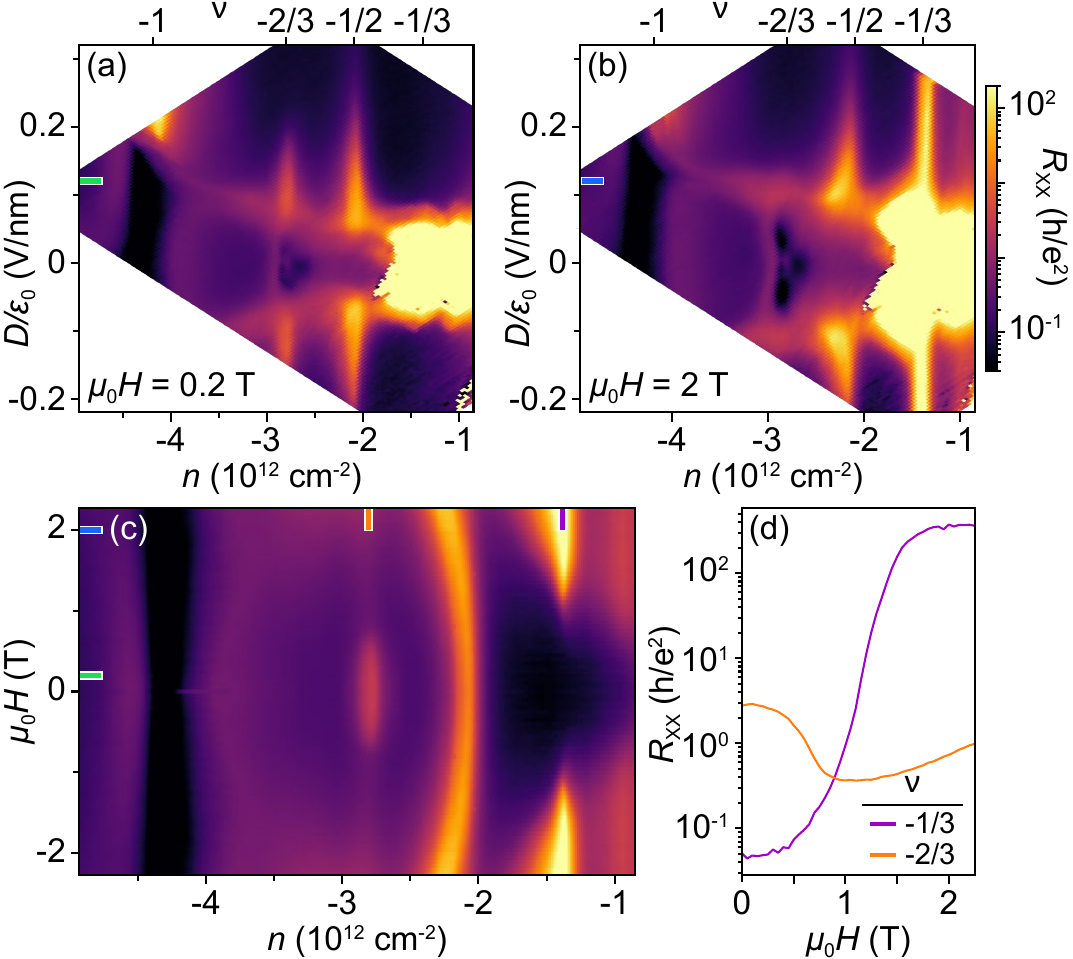}
\caption{
(a) Longitudinal resistance, $R_{\rm XX}$, as a function of carrier density, $n$, and displacement field, $D$, measured at $\mu_0 H = 0.2$~T.
(b) Same as (a) but taken at $\mu_0 H = 2$~T,
(c) $R_{\rm XX}$ as a function of $n$ and out-of-plane magnetic field, $\mu_0H$, measured at fixed $D/\varepsilon_0 = 0.12$~V/nm, showing the evolution of charge-ordered states at commensurate fillings.
(d) Line cuts of $R_{\rm XX}$ at $\nu=-2/3$ and $-1/3$, showing the opposite evolution of the two states relative to $h/e^2$.
}
\label{fig:2}
\end{figure}

At large displacement field, several resistance maxima appear at commensurate fractional fillings. 
Comparing $R_{\rm XX}(n,D)$ at $\mu_0H=0.2$ and $2$~T reveals opposite magnetotransport evolution of the states near $\nu=-2/3$ and $-1/3$; see Figs.~\ref{fig:2}(a) and \ref{fig:2}(b). 
A clear peak at low field near $\nu=-2/3$ becomes strongly suppressed, whereas a barely visible state at $\nu=-1/3$ develops into a narrow resistance peak at high field.
This contrast is resolved continuously in $R_{\rm XX}(n,H)$ at fixed $D/\varepsilon_0=0.12$~V/nm; see Fig.~\ref{fig:2}(c). 
The state at $\nu=-2/3$ is most pronounced at low magnetic field and is progressively suppressed with increasing $H$, where $R_{\rm XX}$ decreases below $h/e^2$, consistent with a crossover toward metal-like behavior; see Fig.~\ref{fig:2}(d). 
By contrast, the state at $\nu=-1/3$ becomes increasingly pronounced and develops into a sharp insulator-like peak above $\mu_0H \sim 1$~T, where $R_{\rm XX}$ increases well above $h/e^2$. 
The state near $\nu=-1/2$ follows a qualitatively different field evolution, addressed below.
The opposite responses at $\nu =- 2/3$ and $-1/3$ persist across four probe configurations in the same device. 
Furthermore, the fractional states are comparatively insensitive to in-plane magnetic fields, at least up to $\sim$1~T, suggesting a strong spin-orbit coupling that favors an out-of-plane spin quantization axis (for more details, see Supplementary Material~\cite{Supplement}).
The contrasting magnetic-field evolution of the two fractional fillings suggests that, although both are associated with commensurate charge order on the triangular lattice, their spin structure is fundamentally different. 

To understand this contrasting response, we consider the generalized Wigner-crystal configurations expected at the two fillings. 
At $\nu=-2/3$, occupation of two of the three triangular sublattices produces a connected honeycomb network, whereas at $\nu=-1/3$, occupation of a single sublattice produces a more dilute triangular network. 
These charge patterns have been observed by imaging in related moiré systems~\cite{Li2021, Li2024}. Their different connectivity naturally gives rise to distinct exchange pathways and magnetic energy scales~\cite{MoralesDuran2022, Kumar2025}.
In particular, nearest-neighbor superexchange on the honeycomb network tends to favor antiferromagnetic correlations, while direct and nonlocal exchange may compete with this tendency~\cite{Tang2023, MoralesDuran2022, Biborski2025}. The complementary charge patterns at $\nu=-2/3$ and $-1/3$ therefore need not exhibit similar magnetic transport.

At $\nu=-2/3$, a perpendicular magnetic field that polarizes the moments competes with antiferromagnetic correlations on the honeycomb network and can modify either the stability of the charge-ordered state or the transport of its charged excitations. 
A reduction of the corresponding activation energy or constraints on charge motion could therefore suppress the resistance while the underlying charge order remains intact, since the charged excitations need not share the magnetic response of the ordered background~\cite{Li2024, Slagle2020}. 
At $\nu=-1/3$, the more dilute triangular network does not provide the same nearest-neighbor exchange pathway; here, spin polarization may instead favor a more localized, high-resistance state~\cite{Spivak2002}. 
This opposite response points to a different balance between spin polarization, charge order, and carrier localization at the two fillings.

\begin{figure}[t]
    \includegraphics[width=1\linewidth]{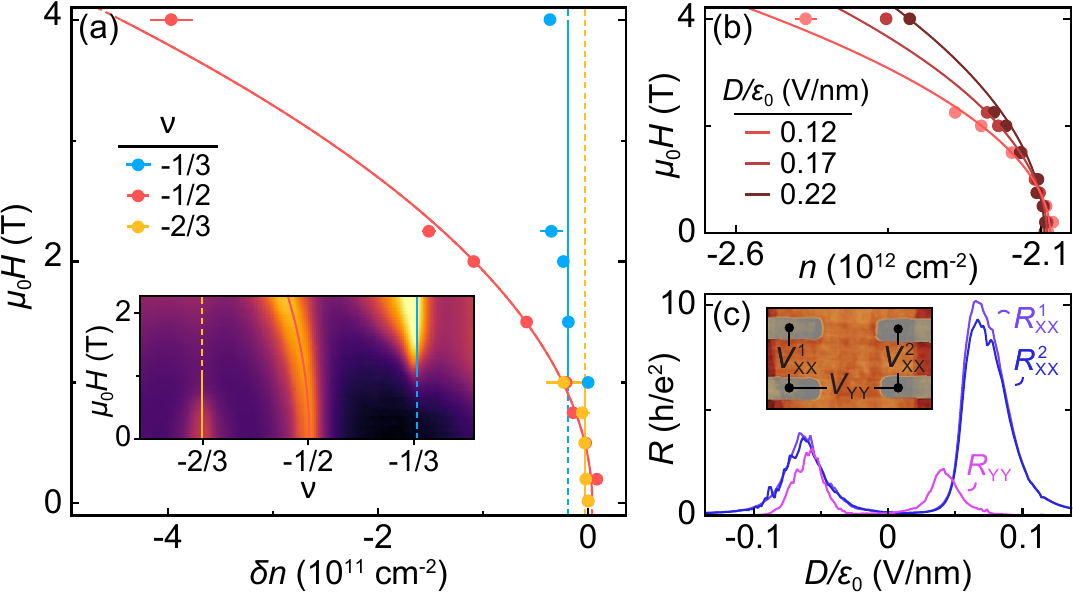}
    \caption{
    (a) Magnetic-field dependence of the extracted relative density shift, $\delta n$, of the resistance peaks at $\nu=-2/3$, $-1/2$, and $-1/3$. 
    The $\nu=-1/2$ data are best described by a shifted quadratic function, whereas the $\nu=-2/3$ and $-1/3$ data are consistent with constants.
    Inset: High-resolution $\nu$--$\mu_0H$ map with the fitted peak positions overlaid.
    (b)~Peak carrier density of the $\nu=-1/2$ state versus $H$ at three displacement fields, showing that the curvature decreases with increasing $D$.
    (c)~Symmetrized longitudinal resistance, R, as a function of $D$, measured near $\nu=-1/2$ in a van der Pauw configuration for three probe configurations shown in the inset. $R_{\rm YY}$ shows a quantitatively different behavior compared to the other two $R_{\rm XX}$ measurements.
    }
    \label{fig:3}
\end{figure}

Finally, we address the state near $\nu=-1/2$ that behaves qualitatively differently. 
While the resistance maxima at $\nu=-2/3$ and $-1/3$ remain near their commensurate densities, the half-filled state shifts systematically with perpendicular magnetic field; see Fig.~\ref{fig:3}(a). Its position follows an approximately quadratic dependence on $H$ over the measured range, in contrast to the linear density dispersion expected for a conventional gapped Chern state~\cite{Polshyn2022}. 
The curvature decreases with increasing displacement field [Fig.~\ref{fig:3}(b)], indicating that the effect weakens deeper in the layer-polarized regime.

A further distinction appears in the in-plane transport. Near $\nu=-1/2$, longitudinal resistance measured along different directions shows pronounced anisotropy [Fig.~\ref{fig:3}(c)]. 
Such behavior is consistent with stripe-like charge order, which breaks the threefold rotational symmetry of the triangular lattice~\cite{Huang2021,Jin2021}. Defects and domain walls in such a state can accommodate changes in density while preserving local stripe correlations~\cite{Matty2022,Tan2023}, providing a possible connection to the field-dependent shift of the resistance maximum. The microscopic origin of both effects, however, remains open.

In summary, we have investigated the magnetic-field evolution of fractional charge-ordered states in the layer-polarized regime of twisted MoTe$_2$. 
The commensurate states at $\nu=-2/3$ and $-1/3$ show opposite responses to perpendicular magnetic field, evolving toward metal-like and insulator-like transport, respectively.
We interpret this contrast in terms of generalized Wigner crystals whose different charge configurations define distinct magnetic lattices and exchange pathways.
Additionally, we identify a qualitatively different state near $\nu=-1/2$ that disperses quadratically with magnetic field and exhibits pronounced transport anisotropy consistent with stripe-like charge order.
The microscopic origin of these features remains an open question. Together, these results reveal a strong dependence of magnetotransport on fractional filling and highlight the interplay between magnetism and charge order in moiré materials.\\

We thank Charles M. Marcus for discussions. We acknowledge support from research grants (Projects No. 43951 and No. 53097) from VILLUM FONDEN, the Danish National Research Foundation, and the European Research Council (Grant Agreement No. 856526).

\bibliography{bibfile.bib}

\end{document}


\title{Supplemental Material:\\
Magnetic-field response of generalized Wigner crystals in twisted MoTe$_2$}
\author{F. K. Wolff}
\affiliation{Niels Bohr Institute, University of Copenhagen, 2100 Copenhagen, Denmark}
\author{S. Jiao}
\affiliation{Niels Bohr Institute, University of Copenhagen, 2100 Copenhagen, Denmark}
\author{H. Park}
\affiliation{Department of Physics, University of Washington, Seattle, Washington 98195, USA}
\author{J. Cai}
\affiliation{Department of Physics, University of Washington, Seattle, Washington 98195, USA}
\author{E. Anderson}
\affiliation{Department of Physics, University of Washington, Seattle, Washington 98195, USA}
\author{X. Liu}
\affiliation{Department of Materials Science and Engineering, University of Washington, Seattle, Washington 98195, USA}
\author{C. Wang}
\affiliation{Department of Materials Science and Engineering, University of Washington, Seattle, Washington 98195, USA}
\author{T.~Taniguchi}
\affiliation{Research Center for Materials Nanoarchitectonics, National Institute for Materials Science, 1-1 Namiki, Tsukuba 305-0044, Japan}
\author{K.~Watanabe}
\affiliation{Research Center for Electronic and Optical Materials, National Institute for Materials Science, 1-1 Namiki, Tsukuba 305-0044, Japan}
\author{X. Xu}
\affiliation{Department of Physics, University of Washington, Seattle, Washington 98195, USA}
\affiliation{Department of Materials Science and Engineering, University of Washington, Seattle, Washington 98195, USA}
\author{D. Xiao}
\affiliation{Department of Materials Science and Engineering, University of Washington, Seattle, Washington 98195, USA}
\affiliation{Department of Physics, University of Washington, Seattle, Washington 98195, USA}
\author{S. Vaitiek\.{e}nas}
\affiliation{Niels Bohr Institute, University of Copenhagen, 2100 Copenhagen, Denmark}

\date{\today}

\maketitle

\section{Sample Fabrication}
The device was fabricated following the same procedure described in \cite{Park2023}. The heterostructure was assembled from mechanically exfoliated flakes of hBN, 2H-MoTe$_2$, and graphite, obtained by standard exfoliation onto a p-doped Si/SiO$_2$ substrate. Suitable hBN flakes for top, bottom, and contact gate dielectrics, as well as thin graphite flakes for gate electrodes, were identified by optical microscopy, and flake quality was confirmed by atomic force microscopy (AFM) or contrast-enhanced optical microscopy.

First the bottom-gate stack was assembled using a polycarbonate-based dry transfer technique, in which a bottom-gate hBN dielectric was picked up first, followed by a graphite bottom-gate electrode, before depositing the stack onto the silicon substrate. Hall bar contacts were defined by standard e-beam lithography, and Pt (8~nm) electrodes were deposited by e-beam evaporation, after which the bottom-gate structure was cleaned by contact-mode AFM. After cleaning the structure was transfered into a glovebox with O$_2$ and H$_2$O levels below 0.1~ppm.

Inside the glovebox, a monolayer MoTe$_2$ flake was exfoliated and cut using an AFM tip. The twisted bilayer was formed by sequentially picking up one half of the monolayer with an hBN flake, rotating the stage by the target twist angle of $3.9^\circ$, and picking up the remaining half. The resulting twisted MoTe$_2$ stack was deposited onto the bottom-gate structure, and residual stamp polymer was removed by sequential washing with molecular sieve-cleaned anhydrous chloroform and dichloromethane. Following encapsulation and polymer removal, the device was taken out of the glovebox and further cleaned by contact-mode AFM.

In order to reduce contact resistance, contact gates were deposited to enable a heavily hole-doped region in the contact areas with a carrier density of approximately $10^{13}$~cm$^{-2}$. The gates were defined by e-beam lithography and Pt (8~nm) evaporation. Cr/Au (5/70~nm) interconnects and bonding pads were added in a subsequent lithography and evaporation step to enable wire bonding. A final round of contact-mode AFM cleaning was performed before completing the encapsulation with a graphite/hBN top gate, whose transfer polymer was removed by chloroform washing in ambient conditions.

\section{Measurement Setup}
Measurements were performed in Oxford Instruments Triton dilution refrigerator with a base temperature of 20~mK using a current-bias configuration by sourcing a current of 1.5~nA through a 100~M$\Omega$ bias resistor. Voltage sourcing and demodulation were carried out using Stanford Research Instruments SR830 and SR860 lock-in amplifiers. 
The current signal was converted to a voltage using a Basel Instruments SP983 $I$--$V$ converter, and voltage differences were amplified by Basel Instruments SP1004 differential amplifiers. The vector magnet has a (1-1-6)~T field range along its three axes.

\begin{figure}[b]
    \includegraphics[width=\linewidth]{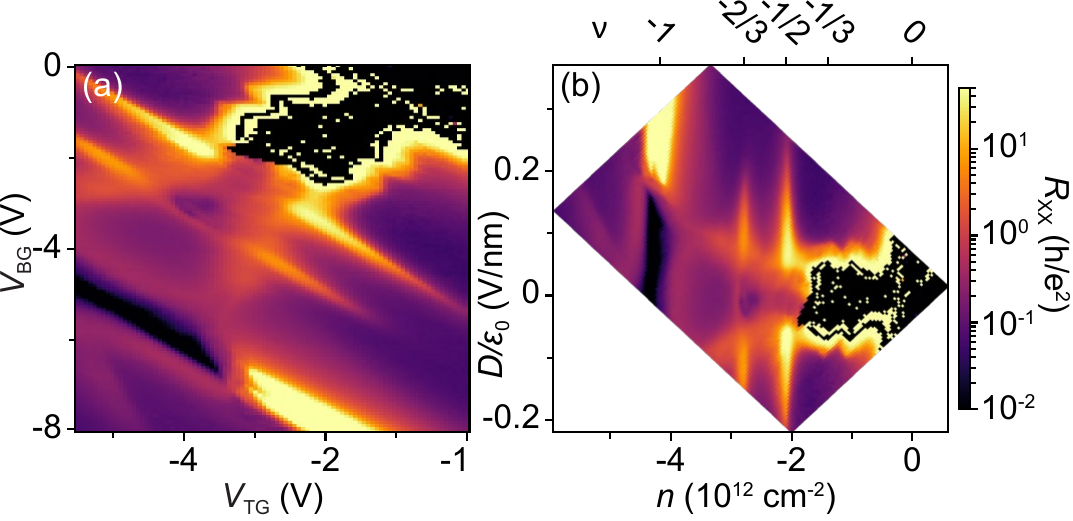}
    \caption{(a) Longitudinal resistance, $R_{\rm XX}$, as a function of bottom-gate voltage, $V_{\mathrm{BG}}$, and top-gate voltage, $V_{\mathrm{TG}}$. The black contour highlights the region used for the conversion to density and displacement field. 
    (b) The same data replotted as a function of carrier density, $n$, and displacement field, $D/\varepsilon_0$. The commensurate states at $\nu=-2/3$, $-1/2$, and $-1/3$ appear as narrow resistance maxima in the layer-polarized regime.
    }
    \label{fig:s1}
\end{figure}

\section{Conversion to $n$--$D$ maps}

The measurements in this work were performed by tuning the top- and bottom-gate voltages, $V_{\mathrm{TG}}$ and $V_{\mathrm{BG}}$. 
To interpret the phase space in terms of physical control parameters, the data were converted from the gate-voltage basis to carrier density, $n$, and perpendicular displacement field, $D$ [Fig.~\ref{fig:s1}]. 
Following the standard parallel-plate-capacitor model for dual-gated moir\'e devices, the two gates independently contribute charge to the twisted MoTe$_2$ bilayer and generate an electric field across the stack. In this approximation, the hole density is given by $n = (C_t V_t + C_b V_b)/e - n_{\mathrm{offset}}$, while the displacement field is $D/\varepsilon_0 = (C_t V_t - C_b V_b)/(2\varepsilon_0) - D_{\mathrm{offset}}$, where $C_t$ and $C_b$ are the capacitances per unit area of the top and bottom hBN dielectrics, respectively, and $n_{\mathrm{offset}}$ and $D_{\mathrm{offset}}$ account for residual offsets. 
Previous optical measurements have established correlated states at the integer and fractional fillings of the moiré bands~\cite{Anderson2023}. We use the features at $\nu=-2/3$ and $-1/2$ as density markers to determine $n_{\mathrm{offset}} = 3 \,(n_{\nu=-2/3} - n_{\nu=-1/2}) - n_{\nu=-1/2}$. The displacement-field offset, $D_{\rm offset}$, is chosen such that the phase diagram is symmetric about $D/\epsilon_0=0$.
This transformation makes the phase diagram easier to interpret, since the relevant low-energy physics is controlled directly by carrier density and interlayer electric field.

Fig.~\ref{fig:s1}(a) shows the measured $R_{\rm XX}$ in the $(V_{\mathrm{BG}},V_{\mathrm{TG}})$ plane. The same data are replotted in the $(n,D)$ basis in Fig.~\ref{fig:s1}(b). In this representation, the commensurate states near $\nu=-2/3$, $-1/2$, and $-1/3$ become directly visible as narrow resistance maxima in the layer-polarized regime, and the broader wedge-shaped crossover between the layer-hybridized and layer-polarized regimes is more clearly resolved. The use of the $(n, D)$ basis throughout the main text and supplement therefore allows direct comparison between electrostatic tuning and the correlated states of the moir\'e lattice.
 
The calibrated moir\'e density also provides an independent estimate of the twist angle. In the small-angle limit,
\begin{equation}
    \theta \approx a\,\sqrt{\frac{\sqrt{3}\,n_{\nu}}{2 \nu}}
\end{equation}
where $a$ is the MoTe$2$ lattice constant~\cite{Reddy2023}. 
Using $a = 3.52$~\AA, $\nu=-1/2$ and $n=-2.1\times10^{12}$~$cm^{-2}$, we obtain $\theta\simeq3.85^\circ$, consistent with the target twist angle of $3.9^\circ$.

\begin{figure}[b]
    \includegraphics[width=\linewidth]{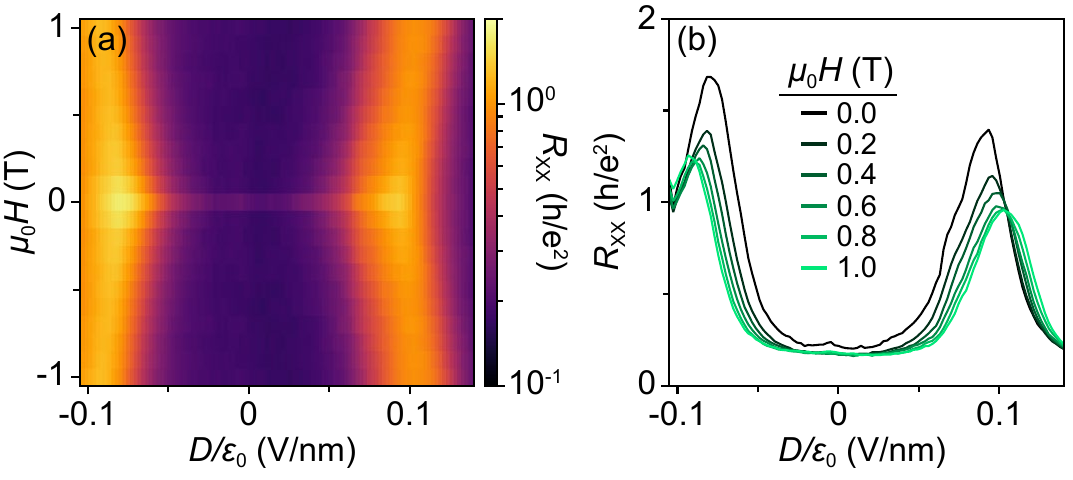}
    \caption{(a) Longitudinal resistance, $R_{\rm XX}$, as a function of displacement field, $D$, and out-of-plane magnetic field, $\mu_0 H$, measured at fixed carrier density $n=-3.7\times10^{12}\,\mathrm{cm^{-2}}$, close to a filling of $\nu\approx -1$. 
    (b) Line cuts extracted from (a) at selected values of $\mu_0 H$. The resistive edge separating the layer-hybridized and layer-polarized regimes shifts to higher $D$ values with increasing $H$.
    }
    \label{fig:s2}
\end{figure}

\begin{figure*}[t]
    \includegraphics[width=1\linewidth]{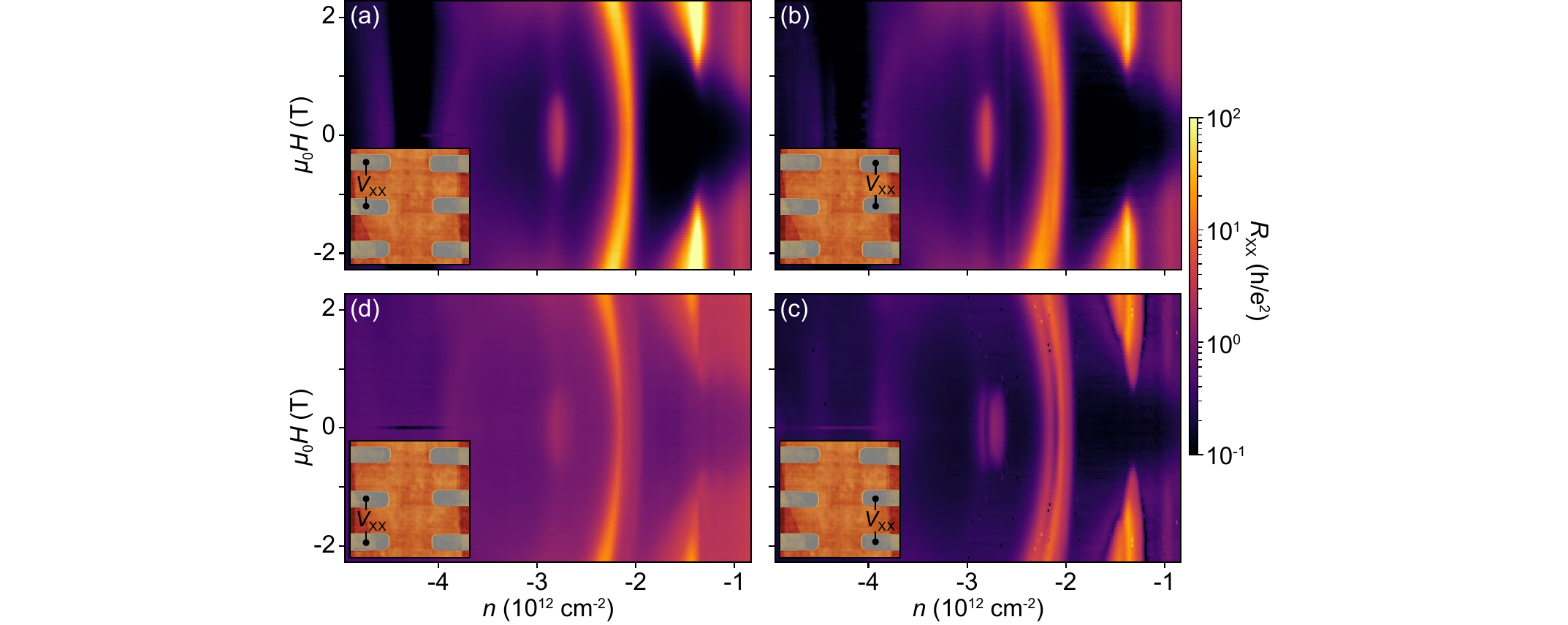}
    \caption{$R_{\rm XX}$, as a function of carrier density, $n$, and out-of-plane magnetic field, $\mu_0H$, measured at fixed $D/\varepsilon_0=0.12 V/nm$ for four probe configurations, indicated by the insets. Here $R_{\rm XX}$ is the magnitude of the measured lock-in signal calculated as $X/cos(\theta)$, where $X$ is the in-phase component of the signal and $\theta$ is its phase.
    Panel (a) shows the same data as in the main-text Fig.~2(c).
    In all configurations, the intensity of the states at $\nu=-2/3$ and $\nu=-1/3$ shows opposite magnetic-field evolution, while the feature near $\nu=-1/2$ remains visible throughout the measured range. The persistence of these trends across probe geometries confirms that the observed field dependence is intrinsic to the device rather than an artifact of a particular contact configuration.
    }
    \label{fig:s3}
\end{figure*}

\section{$H$ dependence of $D_{\rm C}$}

As the out-of-plane magnetic field $\mu_0 H$ is increased, the resistive edge at $D=D_{\rm C}$ separating the layer-hybridized and layer-polarized regimes moves upward in $D$ [Fig.~\ref{fig:s2}].
The origin of this shift is not the main focus of this work, but experimentally it provides additional context for the field evolution of the commensurate states. 
As the boundary moves to higher $D$, a fixed displacement-field cut probes slightly different proximity to the crossover at different magnetic fields. 
The data in Fig.~\ref{fig:s2} therefore confirm that $H$ affects not only the commensurate states themselves, but also the electrostatic boundary between the two underlying lattice regimes.
A possible interpretation of the approximately linear shift is that the layer-hybridized and layer-polarized states have different magnetic polarizations. 
An out-of-plane field would then favor the more strongly polarized state through Zeeman coupling, whereas the displacement field favors layer polarization. 
Since both contributions are approximately linear in $H$ and $D$, respectively, their competition can produce an approximately linear shift of $D_{\rm C}$, provided the polarization difference depends only weakly on $D$.

\section{Signal symmetrization and reproducibility}

The finite size and imperfect alignment of the contacts introduce mixing between longitudinal and transverse voltage drops~\cite{Look1989}. 
To reduce this contribution, the longitudinal resistance, $R_{\rm XX}$, was symmetrized in the out-of-plane magnetic field, $H$, whereas the transverse resistance, $R_{\rm XY}$, was antisymmetrized.
This procedure removes the leading contamination from contact misalignment because the longitudinal and Hall voltages have opposite symmetry under field reversal. 
Measurements were performed with an ac excitation at zero dc bias. 
The field symmetrization (antisymmetrization) therefore provide the relevant correction for extracting $R_{\rm XX}$ ($R_{\rm XY}$) from the measured voltages.

To verify that the magnetic-field evolution of the commensurate states is not an artifact of a particular contact pair, we measured the phase space in four probe geometries [Fig.~\ref{fig:s3}]. The different contact configurations yield somewhat different background resistances, most likely due to poor contact transparency or missing material near the contact. Nevertheless, the correlated features themselves are robust. In all four configurations, the intensity of the states near $\nu=-2/3$ and $\nu=-1/3$ shows opposite magnetic-field evolution, while the feature near $\nu=-1/2$ remains visible throughout the measured field range. This reproducibility indicates that the field dependence discussed in the main text is intrinsic to the device and not due to local contact artifacts.

\section{In-plane magnetic field dependence}

\begin{figure}[t]
    \includegraphics[width=\linewidth]{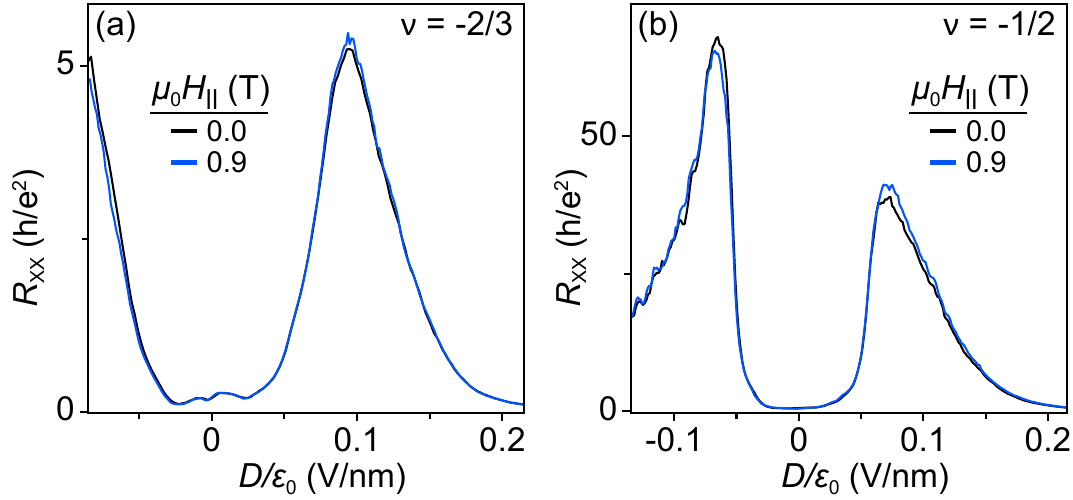}
    \caption{Longitudinal resistance, $R_{\rm XX}$, as a function of displacement field, $D$, measured at fixed carrier density near (a) $\nu=-2/3$ and (b) $\nu=-1/2$, for two values of in-plane magnetic field, $\mu_0H_{\parallel}=0$ and 0.9~T. In both cases, the line shapes show effectively no change with in-plane field, consistent with the weak coupling of the commensurate states to $H_{\parallel}$.
    }
    \label{fig:s4}
\end{figure}

To complete the magnetic-field characterization, we investigated the response to in-plane magnetic field, $H_\parallel$. 
Fig.~\ref{fig:s4} compares displacement-field sweeps taken near $\nu=-2/3$ and $-1/2$ for $\mu_0 H_{\parallel}=0$ and $0.9$~T. Within the experimental resolution, the line shapes are essentially unchanged. No significant shifts or suppression of the commensurate features are observed.
This weak in-plane-field response is consistent with the strong out-of-plane spin locking expected in MoTe$_2$ due to spin--orbit coupling. Together, these data support the interpretation that the dominant field response of the commensurate states is specific to the out-of-plane component of the magnetic field.

\section{Reentrant behavior in temperature}

\begin{figure}[t]
    \includegraphics[width=\linewidth]{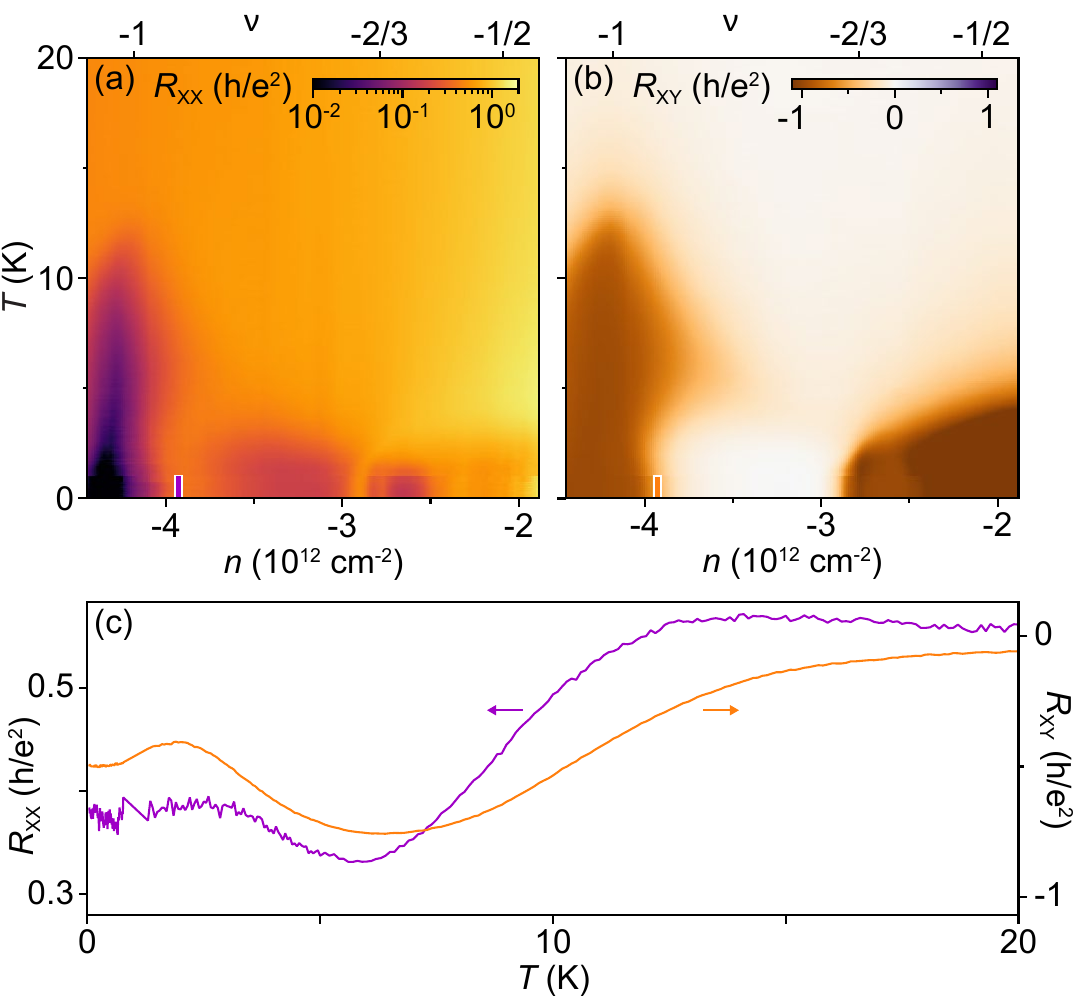}
    \caption{(a) Longitudinal resistance, $R_{\rm XX}$, as a function of carrier density, $n$, and temperature, $T$. 
    (b) Corresponding transverse resistance, $R_{\rm XY}$, over the same range. 
    (c) Representative temperature traces of $R_{\rm XX}$ and $R_{\rm XY}$ at fixed density, just above $\nu=-1$, showing the evolution of the transport response upon warming. The commensurate states weaken with increasing temperature, consistent with their correlated origin.
    All data were taken at $D/\varepsilon_0=0.005$~V/nm and $\mu_0 H=200$~mT.
    $R_{\rm XX}$ and $R_{\rm XY}$ were obtained by symmetrizing and antisymmetrizing the measurements at opposite field polarities, respectively.
    }
    \label{fig:s5}
\end{figure}

As the temperature is raised, the commensurate features near $\nu=-1$, $-2/3$, and $-1/2$ weaken, consistent with their correlated origin [Figs.~\ref{fig:s5}(a) and~\ref{fig:s5}(b)]. 
The temperature dependence is, however, not purely monotonic near $\nu=-1$. 
In particular, the $\nu=-1$ state develops a reentrant extension toward lower density, visible as a shoulder-like feature in $R_{\rm XX}$ and accompanied by a corresponding dip in $R_{\rm XY}$. 
The line-cuts taken in the vicinity of $\nu=-1$ state illustrate this non-monotonic evolution further [Fig.~\ref{fig:s5}(c)].
Although the focus of the present work is the large-$D$ triangular-lattice regime, the reentrant feature near $\nu=-1$ provides useful context for the neighboring correlated states and highlights that multiple competing energy scales are already present in the same device.

\begin{figure}[t]
    \includegraphics[width=\linewidth]{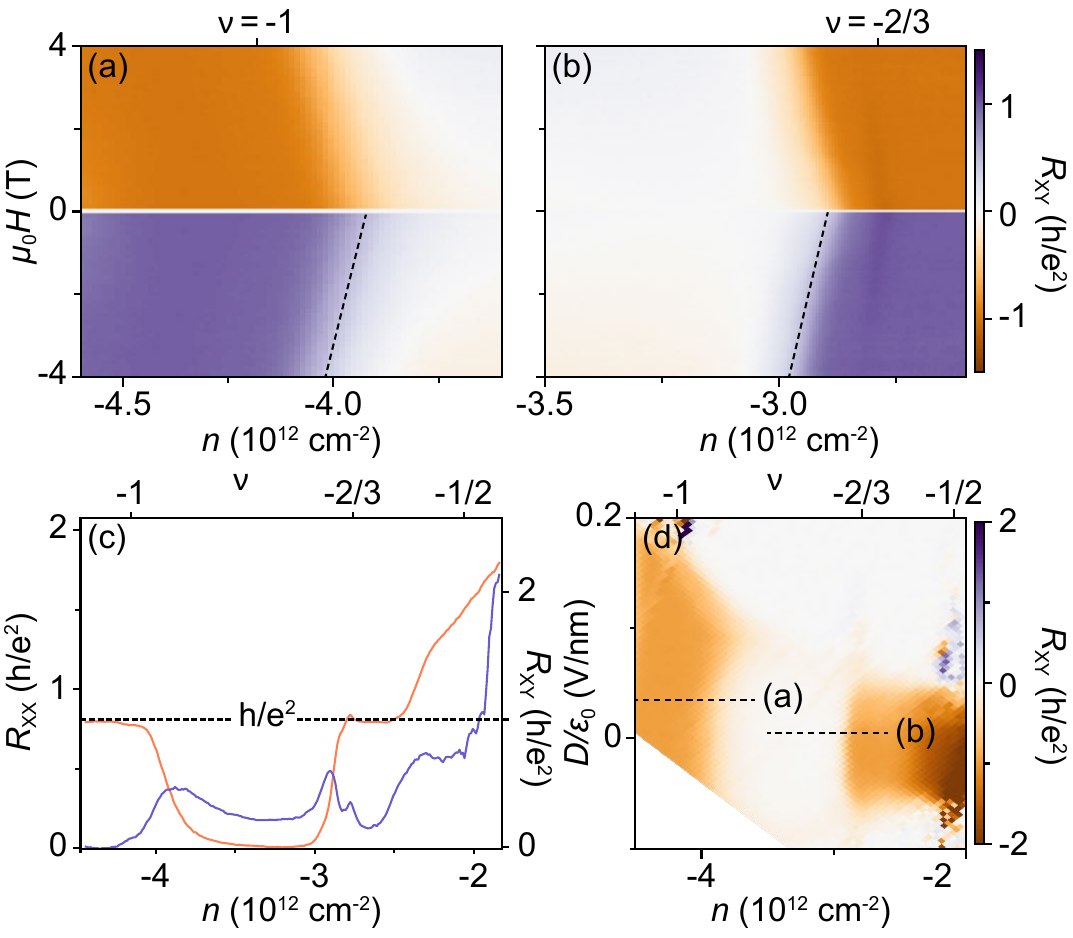}
    \caption{(a) and (b) Transverse resistance, $R_{\rm XY}$, as a function of carrier density, $n$, and out-of-plane magnetic field, $\mu_0H$, measured near (a) $\nu=-1$ at $D/\varepsilon_0=0.035$~V/nm and (b) $\nu=-2/3$ at $D/\varepsilon_0=0.005$~V/nm. The dashed lines indicate the evolution of the corresponding resistance features with field.
    (c) Line cuts of longitudinal ($R_{\rm XX}$) and transverse ($R_{\rm XY}$) resistances as a function of $n$, taken at $D/\varepsilon_0=0.005$~V/nm, showing a quantized plateau near $h/e^2$ for both $\nu=-1$ and $-2/3$.
    (d) Map of $R_{\rm XY}$ as a function of $n$ and $D$.
    $R_{\rm XX}$ and $R_{\rm XY}$ were obtained by symmetrizing and antisymmetrizing measurements at $\mu_0H=\pm 200$~mT, respectively.
    }
    \label{fig:s6}
\end{figure}

\section{Additional Hall-response data}

The Hall response provides additional context for separating the large-$D$ charge-ordered regime from the low-$D$ anomalous Hall regime. 
Near $\nu=-1$ and $\nu=-2/3$, the features in transverse resistance, $R_{\rm XY}$, shift linearly with out-of-plane magnetic field, $H$, with comparable slopes in the two cases [Figs.~\ref{fig:s6}(a) and~\ref{fig:s6}(b)]. 
In both density ranges, the corresponding plateau value is close to $h/e^2$ [Fig.~\ref{fig:s6}(c)].
While this is expected for the $\nu=-1$ state, the analogous value near $\nu=-2/3$ is unusual and remains not fully understood.
A possible scenario for $\nu=-2/3$ is a state emerging from the competition between charge order and anomalous Hall character, such as the QAH crystal proposed in Ref.~\cite{Song2024}, though incomplete equilibration of edge modes may also play a role~\cite{Grivnin2014}.

This behavior differs qualitatively from the large-$D$ layer-polarized regime studied in the main text. There, the state near $\nu=-2/3$ evolves into a topologically trivial charge-ordered insulator whose longitudinal resistance is strongly suppressed by $H$, rather than following the approximately linear Hall-feature evolution observed near zero displacement field. The Hall data in Fig.~\ref{fig:s6} therefore help distinguish the low-$D$ anomalous Hall regime from the large-$D$ commensurate charge-ordered regime.

\section{Theoretical model}

In the layer-polarized regime, the low-energy degrees of freedom can be described, to leading order, by carriers on an effective triangular lattice. 
At the commensurate fillings $\nu=-2/3$ and $-1/3$, interactions can stabilize $\sqrt{3}\times\sqrt{3}$ charge order~\cite{Regan2020, Tang2020, Xu2020, Huang2021}.
We use Hartree–Fock and exact-diagonalization calculations to examine how the favored states evolve between spin-unpolarized and spin-polarized sectors. 
The calculations are intended to provide a qualitative comparison with the contrasting magnetotransport observed experimentally.

\begin{figure}[t]
    \includegraphics[width=\linewidth]{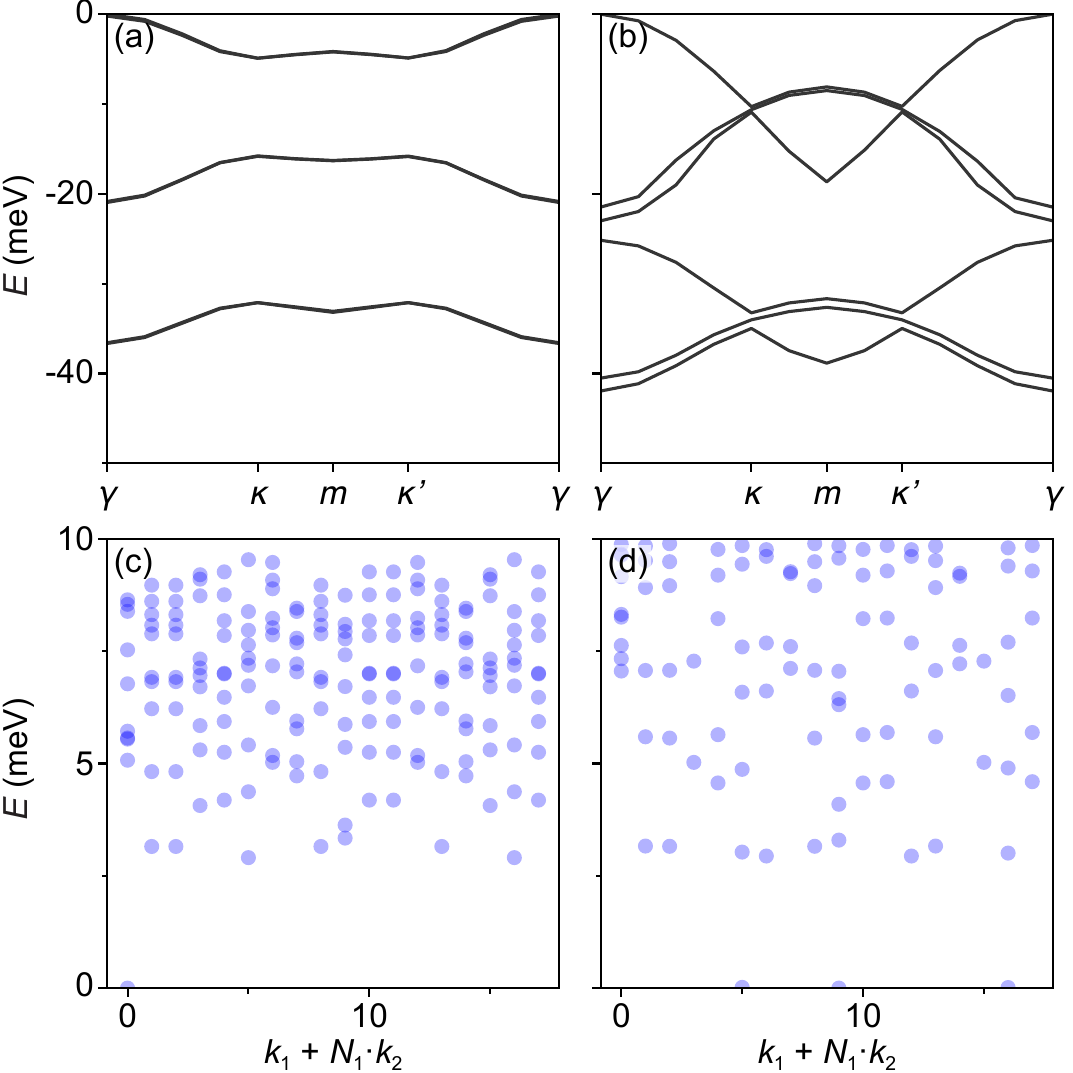}
    \caption{  
    (a) and (b) Hartree-Fock band structures at $\nu=-2/3$ for the spin-unpolarized antiferromagnetic charge-ordered state and the spin-polarized metallic state, respectively.
    (c) and (d) Exact-diagonalization spectra at $\nu=-1/3$ in the spin-unpolarized and spin-polarized sectors, respectively.
    The spin-unpolarized sector has a nondegenerate ground state with no clear signature of $\sqrt{3}\times\sqrt{3}$ charge order, whereas the spin-polarized sector exhibits a threefold-degenerate ground state consistent with $\sqrt{3}\times\sqrt{3}$ charge order.
    Panel (d) is shifted upward relative to (c) by 7 meV.
    }
    \label{fig:s7}
\end{figure}

At $\nu=-2/3$, we use Hartree–Fock calculations with a relative dielectric constant of $\epsilon = 20$ and a total of 48 $k$-points sampled within the first Brillouin zone of the supercell.
We find that the spin-unpolarized sector favors a collinear antiferromagnetic charge-ordered state with an insulating band structure [Fig.~\ref{fig:s7}(a)]. 
In the spin-polarized sector, the favored state becomes metallic [Fig.~\ref{fig:s7}(b)].
This provides a possible microscopic route for the suppression of resistance observed experimentally with increasing perpendicular field. 
The calculation therefore supports the idea that spin polarization can destabilize the low-field insulating state.

At $\nu=-1/3$, Hartree–Fock strongly favors symmetry-broken charge-ordered solutions and does not reproduce the weakly resistive low-field state observed experimentally. 
Because Hartree–Fock can overestimate the stability of symmetry-broken states, we instead examine this filling by exact diagonalization on an 18-site lattice with $\epsilon=8$. 
In the spin-unpolarized sector, the ground state is nondegenerate and shows no clear signature of $\sqrt{3}\times\sqrt{3}$ charge order [Fig.~\ref{fig:s7}(c)]. In the spin-polarized sector, a threefold-degenerate $\sqrt{3}\times\sqrt{3}$ charge-ordered state appears [Fig.~\ref{fig:s7}(d)]. 
Within this picture, spin polarization can favor charge order at $\nu=-1/3$, qualitatively consistent with the emergence of a high-resistance state in magnetic field.


\bibliography{bibfile.bib}

\newpage